# Radius and Chirality Dependent Conformation of Polymer Molecule at Nanotube Interface

*Chenyu Wei*

MS 229-1, NASA Ames Research Center, Moffett Field, California, 94035

cwei@mail.arc.nasa.gov

Temperature dependent conformations of linear polymer molecules adsorbed at carbon nanotube (CNT) interfaces are investigated through molecule dynamics simulations. Model polyethylene (PE) molecules are shown to have selective conformations on CNT surface, controlled by atomic structures of CNT lattice and geometric coiling energy. PE molecules form entropy driven assembly domains, and their preferred wrapping angles around large radius CNT (40, 40) reflect the molecule configurations with energy minimums on a graphite plane. While PE molecules prefer $0^o$ wrapping on small radius armchair CNT (5, 5) predominantly at low temperatures, their configurations are shifted to larger wrapping angle ones on a similar radius zigzag CNT (10, 0). A nematic transformation around 280 K is identified through Landau-deGennes theory, with molecule aligning along tube axis in extended conformations.

Interfaces between polymer molecules and carbon nanotubes (CNTs) are present in many CNT based nanomaterials and applications, such as polymeric nanocomposites [1-6], CNT sensors or functionalization with large organic or biological molecules [7-9] and molecule transports in CNT channels [10, 11]. More recently ordered and selective assemblies of polymer and DNA molecules at nanotube interface have been investigated, such as in PPEI-EI-single wall CNT (SWCNT) thin film [12], at PmPV-SWCNT [13], stearic acid molecule-SWCNT [14], and polyaniline-multiwalled CNT interface [15] systems. DNA molecules have also been shown to adsorb on CNTs with dependence on tube radius and DNA sequence [16]. These studies open opportunities to use CNTs as substrates for molecule self-assemblies. The molecule structures and thus atomic interactions at CNT interfaces are also expected to significantly influence many properties in above mentioned nanosystems. For example molecule structure ordering enhanced interfacial mechanical transfers in CNT composites [17] and conformation dependent diffusivity of biopolymers in CNTs [11] have been shown in recent studies.

In this letter we study conformations of polymer molecules at various CNT interfaces through molecular dynamics (MD) simulations [18]. Model polyethylene (PE) molecules are shown to have selective conformations on CNTs, with strong dependence on temperature and on the radius and chirality of nanotubes. The molecules are found to form assembly domains around a large radius CNT with preferred wrapping angles, which reflect configurations at the interaction energy minimums with the CNT lattice. While wrappings around $0^\circ$ dominate on small radius armchair CNT, molecule wrappings shift to larger angles on similar radius zigzag tube.

PE is chosen as model polymer in our simulations, which provides a good representation for long linear molecules. A typical simulation unit cell is about 30Å×30Å×209Å for initial configurations with periodical boundary conditions. Total of 50 PE molecules with 100 repeating $CH_2$ units and a capped 200Å long CNT (10, 0) is mixed together as a composite material. The material is prepared at high temperature of 600 K with each individual molecule relaxed through a Monte Carlo simulation beforehand, and is gradually cooled down to low temperatures with rate 10 K/100 ps at a constant pressure (P = 1 bar). A united atom model is used for PE intramolecule interactions with bond stretching, bending, and dihedral potentials. A truncated 6-12 Lennard-Jones (LJ) type van der Waals (VDW) interaction is included between CNTs and matrix and within matrix. The details of the force field can be found elsewhere [19]. The Amber force field [20] has been used for the C-C interaction on the CNT. Time step of 0.5 fs is used, and all the data shown in this letter is statistical average over 100 ps MD simulation interval and over eight sample sets.

The radial distribution function (RDF) $g(r)$ of the PE molecules (ones near the CNT cap region are excluded) around CNT (10, 0) as a function of radial distance $r$ from the CNT wall is shown in Fig. 1a, at $T$ = 50 K to 600 K. Two discrete adsorption layers can be observed at 3.5 and 7.5 Å due to the interfacial VDW interaction. Similar layered structures have been observed in polymer melts between flat plates [21, 22] and in short polymer molecules around CNTs [17, 23] in previous simulation studies. The temperature effects on the RDF have two parts: enhanced adsorption at low temperatures; and broaden peaks with positions shifted to higher values at high temperatures, due to the thermal expansion of the composite.

The conformation of the molecules is further investigated through an orientation ordering parameter measuring the correlation between the local orientation of molecule backbones and the CNT axis [17, 24]. The parameter is defined as $S_z(r) = 0.5 \times [3\langle \cos^2 \theta(r) \rangle - 1]$, where $\theta$ is the angle between the vector connecting the two ends of a 4-segment subchain on a molecule and the CNT axis, $Z$; $r$ is the distance of the center mass of the subchain from the CNT wall in the $XY$ plane (see inset of Fig. 1a for definitions); and $\langle \rangle$ represents the statistical average over all the molecules in the simulation. $S_z$ = -0.5, 0, or 1 represents cases for the molecules in perpendicular, random, or aligning direction with the tube axis, respectively. The function $S_z$ is plotted in Fig. 1b at various temperatures, and it can be seen that a higher ordering (larger $S_z$) in molecule orientations is correlated with the higher density in the adsorption layers, within which the molecules prefer extended conformations along the tube axis, induced by gains in the adsorption energy. A recent experiment has shown that micrometer long rod-like fd virus can induce a similar elongation in polymer molecules [25].

With the decrease of temperature the PE molecules rearrange their alignment with the CNT to lower adsorption energy further, as indicated from the enhanced magnitude of the peaks in $S_z$ (Fig. 1b). Following the Landau-deGennes theory on isotropic-nematic phase transition [26], a correlation volume is defined as $V_\xi = 4\pi \int_0^\infty r^2 S_Z(r)g(r)dr$ for molecules in extended conformations along tube axis, which diverges around a transition temperature $T_c$, as $V_\xi \sim T/(T-T_c)$. The $1/V_\xi$ as a function of $1/T$ is shown in inset of Fig. 1b, and an extrapolation value of $T_c \approx 280K$ is obtained at $1/V_\xi = 0$. Different from the RDFs, where the peaks of the molecule density continue to increase with lowering $T$ down to 50 K due to the continuing volume compression of the composite, the magnitude of the peaks in $S_z$ only increases significantly till a temperature around glass transition temperature $T_g$ ($\approx$ 310 K from density vs. temperature function) of the composite. The conformations of the PE molecules begin to freeze and further alignments with the embedded CNT are dramatically slowed when $T < T_g$.

While the ordering parameter $S_z$ is an averaged value, the conformation of an individual molecule, especially the ones at the interface, can be greatly influenced by the atomic structure of the substrate CNT. A previous theoretical analysis has shown that geometry constrains can induce interesting features in coiling angle for wide strapped polymer molecules, though effects of temperature and atomic registry of nanotubes were not considered there [27]. For a PE molecule adsorbed on a graphite plane, there are three energetically favorable configurations in commensuration with the substrate, each separated by $60°$, if bond length and angle in graphite and polymer matches. When mismatch exists, the energy profile for an adsorbed molecule is more complicated. Shown in Fig. 2a is the VDW interaction energy as a function of the rotation angle of a straight (rigid) 100-unit PE molecule (in united atom model) at energy minimization distance on a graphite plane (substrate fixed). The parameters for VDW are same as in the composites. The C-C bond length is $1.53 Å$ and $1.42 Å$, and C-C-C bond angle is $112.8°$ and $120°$ for PE molecule and graphite, respectively. The illustration for the rotation is shown in inset of Fig. 2a. The rotation step is $0.1°$ and the translations in the graphite plane (by $0.142 Å$ per step) are allowed to minimize the VDW energy. Only angles between $0°$ and $60°$ are considered due to symmetry. It can be seen in Fig. 2a that there are five energy minimum regimes around following angles: (1) $0°$ and $3.6° = (120-112.8)°/2$ due to the mismatch in bond angles; (2) $60°$ and $(60-3.6)°$ similarly; (3) $15.3° = \cos^{-1}(a_1/a_2)$, where $a_1 = 2\times\sin(120°/2)\times1.42Å$ on CNT and $a_2 = 2\times\sin(112.8°/2)\times1.53Å$ on graphite (see inset of Fig. 2a) due to mismatch in bond lengths; (4) $44.7° = (60-15.3)°$ similarly; (5) $30°$, due to geometric symmetry. These regimes are separated by flat distribution in between. The reasons for the non-smoothness and fluctuations in the energy profile are attributed to the finite length of the PE molecule and the atomic level (which is discrete) description for the interactions. Similar patterns and minimums have also been observed for CNTs rolling on a graphite plane [28]. For other local minimums such as at $6°$ with narrow distribution, their effects on molecule conformation could be limited.

When the graphite plane rolled into a nanotube, the rotation angles at the VDW interaction energy minimums are projected into favorable wrapping angles for the polymer molecules on CNT (*n, m*), with a shift of $\sin^{-1}(\sqrt{3}m/2\sqrt{n^2+m^2+nm})\pm30°$. Several PE-CNT composite systems are studied with procedures described above, which consists of 200Å long continuous CNT (5, 5), (10, 0), (10, 10), or (40, 40) and 95, 95, 115, or 250 100-unit PE molecules, respectively. At the interface of the large radius CNT (40, 40) the preferred wrapping angles are expected similar as the ones on the graphite plane, as the energy costs for coiling a molecule around the tube is small. Shown in Fig. 2b is a snapshot of atomic structure of PE molecules at the interface of CNT (40, 40) at *T* = 50 K in one of the samples. It can be seen that PE molecules are assembled on the nanotube in domains. Within each domain the molecules are in ordered configurations with a preferred wrapping angle and are separated by equal distance of 4.26Å. Such ordered structures are attributed to the interaction with the CNT lattice, while the interactions between polymer molecules are also helpful for the ordering. The marked values ($\sim 3°, 15°, 26°/33°, 55°, 75°$) in Fig. 2b reflect the angles at minimums in Fig. 2a, though small deviations exist. The reason for the deviations is that the molecules can be tilted by a small angle due to interactions at domain boundaries. Similar feature has been observed for alkane molecule assembly on graphite in previous experiment [29] and simulation [30]. Multiple domains coexist due to gains in entropy, which scales as $S = k_B N_d \ln M_d$, where $N_d$ and $M_d$ is number of possible wrapping angles and domains on a CNT, respectively. For the CNT (40, 40) composite studied here the entropy contribution to the free energy is estimated as $TS \sim 40 k_B T$, if taking $N_d \sim 1/2 \times 360/60 \times 5$ (there are 5 main minimums in Fig. 2a and factor 1/2 is due to symmetry), and $M_d \sim 2 \times 7$ (there are 7 main domains on one side of the CNT in Fig. 2b and factor 2 is for both sides). Such contribution is comparable to the cost in PE adsorption energy $\Delta E \sim 70 k_B T_{300K}$ for $T > 300$ K (assuming half of the 13600 atom sites on the CNT is for adsorption and considering $\Delta E \sim 0.03 eV \sim k_B T_{300K}$ per 100 atoms between $\theta = 3.6°$ and the plat regime in Fig. 2a). While domains can merge to lower the interaction energy at boundaries, the dynamics is expected to be slow, especially below $T_g$, and multi domains can remain at low $T$ due to molecule conformation frozen.

The coiling of the PE molecules around a nanotube is quantitatively described through a probability function $P(\theta)$ for local wrapping angle $\theta$ between the vector connecting the two ends of a 3-segment subchain on a PE molecule and the nanotube axis (see Fig. 1a inset). The normalized $P(\theta)$ with $\int P(\theta) d\theta = 1$ in the first adsorption layer is plotted in Fig. 3 for various PE-CNT interfaces, at $T = 50$ K to 500 K. It can be seen that distinct peaks appear with decreasing temperature, representing the formations of molecule assembly domains. The abundant peaks ($\sim -75°, \pm 60°, -41°, -20°, 0°, 31°$) on CNT (40, 40) (Fig. 3a) through out the wrapping angle range are expected as already shown in Fig. 2b. Reasons for deviations from the rotation angles at energy minimums in Fig. 2a were discussed above, and such derivations also contribute to the broad width of the peaks in $P(\theta)$.

For a much smaller radius CNT such as (5, 5) ($r \sim 3.5$Å) or (10, 0) ($r \sim 3.9$Å) the molecule coiling energy becomes important, which scales as $E_{coil} = k_c \sin^4 \theta / R^2$, where $k_c$, $\theta$, and $R$ is the bending constant, coiling angle and radius for molecules, respectively [31]. Large angle wrapping is much depressed on these tubes as shown in Fig. 3b and 3d. Due the curvature effect on a small radius tube, the molecule wrapping angle $\theta$ is mapped into a projected wrapping angle $\theta_p$ on to the CNT surface (see Fig. 4 for illustration) through following relation, $\tan\theta_p = r_{CNT}/(r_{CNT}+\Delta)\times\tan\theta$, where $r_{CNT}$ is the nanotube radius, and $\Delta \approx 3.5\text{Å}$ is the distance between the CNT and the molecules in the first adsorbed layer. The values for the preferred wrapping angles $\theta_P$ are expected to be modified slightly towards to smaller values compared with on a flat graphite surface due to $E_{coil}$. On the armchair CNT (5, 5) a significant peak around $0°$ dominates as shown in Fig. 3b, which we attribute to the double favorable factors from the minimum VDW interaction energy with CNT lattice and the minimum molecule coiling energy at such configuration (due to the broad width of the peak, wrapping at $\theta = 3.6°$ is difficult to be distinguished from $0°$ in Fig. 3b). The distribution of $P(\theta)$ is much broader on the similar radius zigzag CNT (10, 0), with two peak regimes around $-22.5°$ and $15°$ as shown on Fig. 3d. The wrapping angles $\theta$ of $-22.5°$ and $15°$ represents $\theta_P$ of $-12.5°$ and $8°$ on the CNT (10, 0), respectively. While the former is associated with rotation angle around $-(30-15)°$ at the local energy minimum, the later has rather large derivation from the angles at the energy minimums in Fig. 2a. Angle tilting due to domain boundary interactions and thermal fluctuations can contribute to such deviation. Similarly on a medium radius CNT (10, 10) nanotube, two peak regimes are observed around $\pm 22°$ in Fig. 3c, and their projected wrapping angles onto the CNT are $\pm 15°$, matching two of the minimums in Fig. 2a. Assembly domains with $0°$ wrapping on CNT (10, 10) interface are observed in examining the atomic structures, though the broad peaks at $\pm 15°$ smear the distribution of $P(\theta)$ near $0°$. The larger value of $P(\theta)$ at $\pm 15°$ than $0°$ are contributed to the favorable formation of multi domains and to the finite (8 here) sample sets in calculating $P(\theta)$.

A stick-ball plot of a snapshot of PE molecules at CNT (5, 5) interface (portion of the nanotube in the simulation) is shown in Fig. 5, It can be seen that the molecules are in well extended and ordered structures with wrapping angle of $0°$ at low temperature (T = 50 K, up part of Fig. 5), when the composite is in solid state. A molecule (black sticks) is seen in a clear registry structure with the substrate CNT. At high temperatures above glass transitions, molecules begin to choose orientations away from the nanotube axis and the increased thermal fluctuations begin to destroy the ordered structures, as shown in the stick-ball plot of polymer molecules at the same section of the CNT (5, 5) interface in the sample set, at T = 600 K (bottom of Fig. 5). Similarly shown in Fig. 6 is the snapshot for PE molecules at the surface of CNT (10, 0) (portion of the nanotube in the simulation) at T = 50K. It can be seen that the molecules are in ordered structures with wrapping angle around 22° and −22° (see above discussion) in two adjunct domains.

With increase of temperature, the influence from the atomic level interfacial interactions diminishes, and the conformation of polymer molecules would be dominated by more general geometric feature such as tube radius. Shown in Fig. 7 is the logarithm of $P(\theta)$ as a function of $\sin^4\theta/k_BT$ for various PE-CNT interfaces at $T = 600$ K. Flattening of $\ln P(\theta)$ with increasing tube radius and similarity in distribution for similar radius CNT (5, 5) and (10, 0) can be observed. These features are expected according to Boltzmann distribution $P(\theta) \propto e^{-\langle n_s \rangle l_0 E_{coil}/k_BT} = e^{-\langle n_s \rangle l_0 k_c \sin^4\theta/R^2 k_BT}$, where $\langle n_s \rangle$ is the average number of connecting units on a PE molecule with a same wrapping angle, $l_0 \sim 1.32$Å is length of the C-C bond projected onto the molecule backbone, and $k_c = L_p k_B T$ is bending force constant (persistence length $L_P \sim 10$Å at 300 K for PE molecule [32]). The linear fitting of $\ln P(\theta)$ vs. $\sin^4\theta/k_BT$ in Fig. 7 gives $\langle n_s \rangle \sim 3-4$ for PE molecules on CNT (5, 5), (10, 0), and (10, 10), with $\langle n_s \rangle l_0 \sim 4$Å $\sim L_P$ at $T = 600$ K. A much larger $\langle n_s \rangle \sim 19$ is calculated on CNT (40, 40), indicating favorable extended molecule conformations on the much large tube even at high temperature.

In summary the conformation of linear polymer molecules at a CNT interface is studied through MD simulations, which are shown to strongly depend on the lattice structure of the substrate nanotubes and on the temperature. Entropy driven molecule assembly domains are found, and the preferred wrapping configurations on large radius CNT reflect the ones on a graphite plane with interaction energy minimums. While a PE molecule prefers 0° wrapping around a small radius armchair CNT, it prefers wrappings with larger angles around a similar sized zigzag tube. A nematic transformation is found around 280 K, with molecules in extended conformation along the nanotubes. These results are important not only for understanding the interfacial interactions and structural couplings presented in CNT based composites and sensors, but also for designs of radius and chirality dependent molecular assemblies on CNT surfaces.

**Acknowledgment.** This work is partially supported by NASA contract NAS2-03144 to UARC. We thank T. Yamada for useful comments.

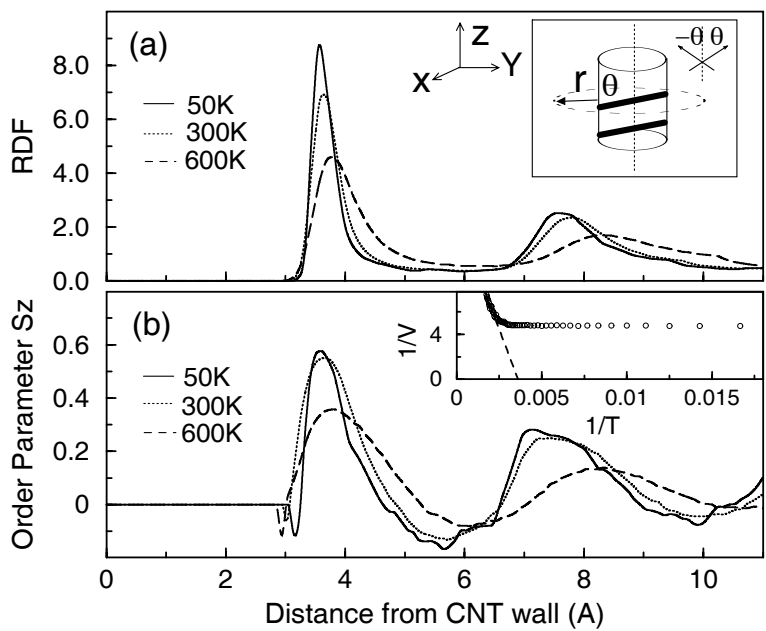

**Figure 1**. (a) The radial distribution function (RDF) as a function of distance from CNT (10, 0) wall for PE molecules. Inset: Schematic plot of an embedded CNT in a simulation unit cell and definition of coordinates. The black strap represents a molecule around the CNT with a wrapping angle $\theta$. The orientation vector of a subchain in a molecule is shown at symmetric angle $\pm\theta$ with the tube axis (up-right). (b) The orientation order parameter $S_z$. Inset: The inverse of correlation volume ($\text{Å}^{-3}$) as a function of $1/T$ ($\text{K}^{-1}$).

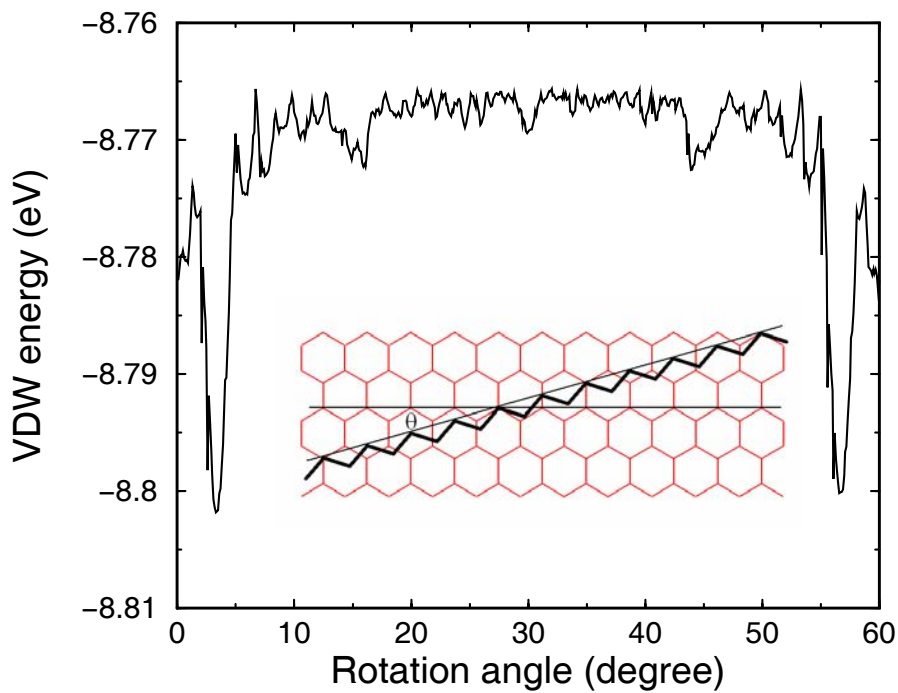

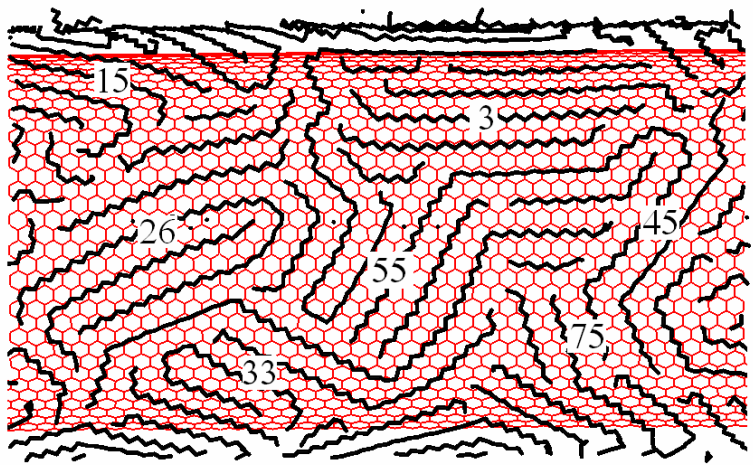

**Figure 2**. (a) The VDW interaction energy between a 100-unit PE molecule and a graphite plane as a function of the rotation angle of the molecule. Insert: Schematic plot for the rotation of the molecule (partially shown). The line running parallel represents the backbone of the molecule at initial configuration with $\theta = 0^\circ$. (b) A snapshot of the atomic structure at the interface of a PE-CNT (40, 40) composite (partially shown, $T = 50$ K.) Values of wrapping angles in each domain are marked.

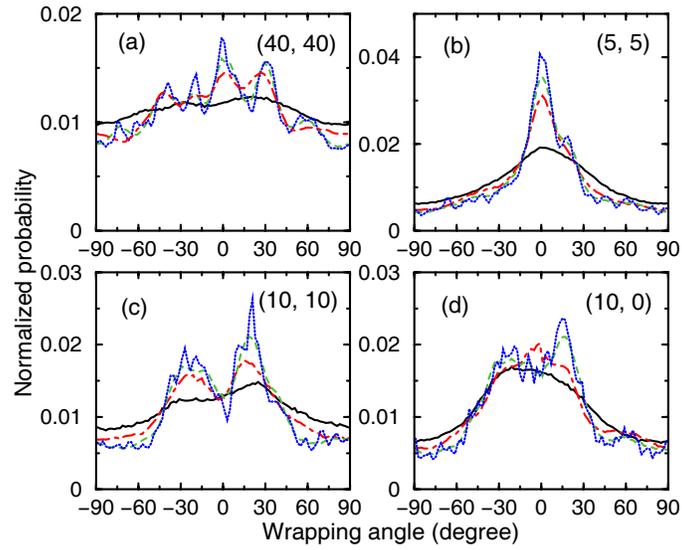

**Figure 3**. The probability function $P(\theta)$ at (a) CNT (40, 40), (b) (5, 5), (c) (10, 10), and (d) (10, 0) interface. Solid, dot-dashed, dashed, and dotted line is for $T$ = 500 K, 400 K, 300 K, and 50 K respectively.

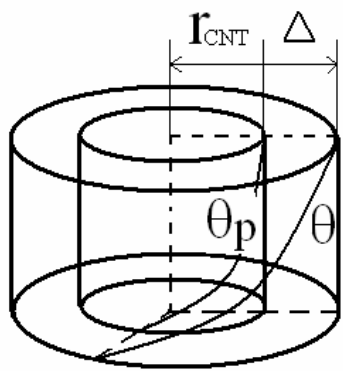

**Figure 4**. Illustration of a polymer molecule (solid curve) adsorbed on CNT. The molecule wrapping angle $\theta$ on the adsorption surface (at distance $\Delta$ from CNT wall) is projected onto the CNT surface with wrapping angle $\theta_p$.

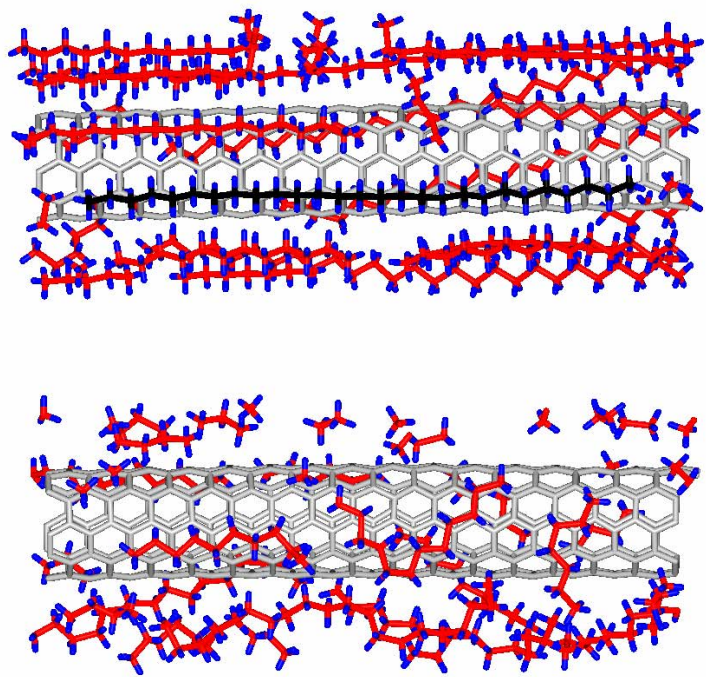

**Figure 5**. Up: Stick-ball plot of a snapshot of polyethylene molecules at the interface of CNT (5, 5) at T = 50 K (portion of the nanotube in simulation is shown). The molecule shown in black is in registry with the CNT lattice. The hydrogen atoms shown in the plot are not explicitly included in simulations. Other molecules in the system are not shown for clarity. Bottom: Molecules at the same section of the nanotube in the same sample, at T = 600 K.

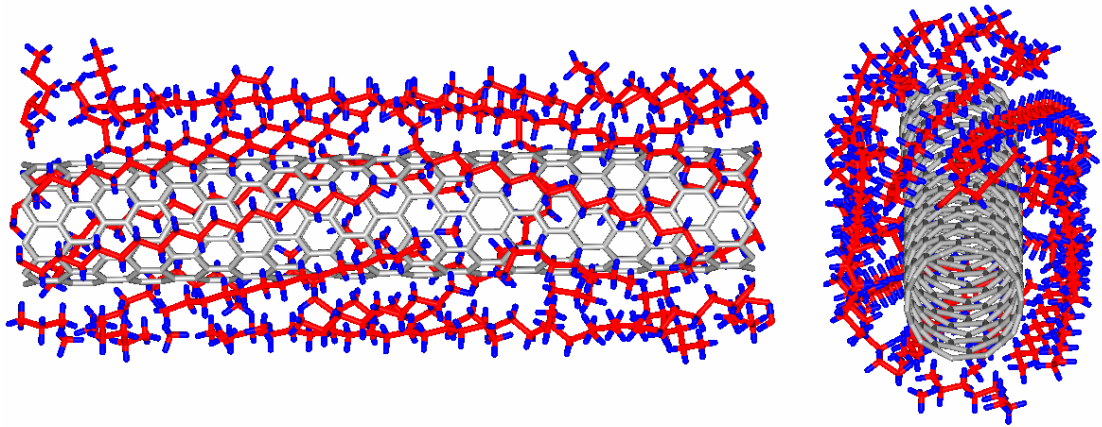

**Figure 6**. Left: Snapshot of polyethylene molecules at the interface of CNT (10, 0) at T = 50 K (portion of the nanotube in simulation is shown). Right: Titled top view.

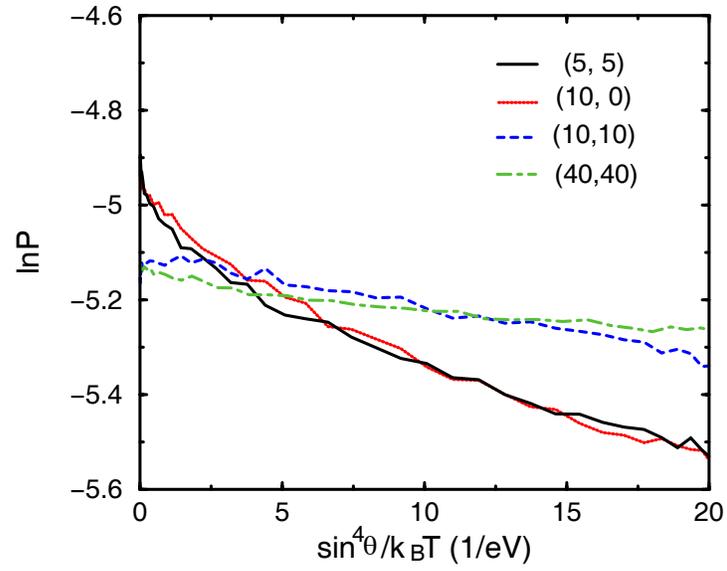

**Figure 7**. The logarithm of $P(\theta)$ as a function of $\sin^4\theta/k_BT$ ($eV^{-1}$) at $T = 600$ K. The solid, dotted, dashed, and dot-dashed line is for CNT (5, 5), (10, 0), (10, 10), and (40, 40) composite, respectively